# Advancing X-Ray Camera Front-End Electronics Architecture through the AXIS-TAP Technology Platform

Jill Juneau*[a], Declan O'Neill [b], Peter Orel [b], Eric Miller [a], Sven Herrmann [b], Gregory Prigozhin [a], Robert Goeke [a], Elio Angile [a], Mark Bautz [a], Steven Allen [b,d,e], Tanmoy Chattopadhyay [b], Kevan Donlon [c], Michelle Gabutti [a], Catherine E Grant [a], Beverly LaMarr [a], Christopher Leitz [c], Glenn Morris [b,d], Abigail Y. Pan [b,d], Tonya Peshel [b,d], Artem Poliszczuk [b], Haley Stueber [b,d], Keith Warner [c]

[a]MIT Kavli Institute for Astrophysics and Space Research, 70 Vassar St, Cambridge, MA 02139;
[b]Kavli Institute for Particle Astrophysics and Cosmology, Stanford University, 452 Lomita Mall, Stanford, CA 94305, USA;
[c]MIT Lincoln Laboratory, 244 Wood St building 1324, Lexington, MA 02421, USA
[d]Department of Physics, Stanford University, 382 Via Pueblo Mall, Stanford CA 94305, USA
[e]SLAC National Accelerator Laboratory, 2575 Sand Hill Road, Menlo Park, CA 94025, USA
*jjuneau@mit.edu

## ABSTRACT

Advancing time-domain soft X-ray astronomy demands focal plane readout electronics capable of high-speed digitization with low noise while maintaining spectral fidelity across multiple detector channels. The Advanced X-ray Imaging Satellite (AXIS), a NASA Probe-class mission concept, provided the driving requirements and heritage for developing a generalized high-speed (≥5 fps) camera Front-End Electronics (FEE) architecture applicable to future soft X-ray missions. The FEE architecture defines a modular, dual-box redundant configuration to operate a focal plane of four 16-channel CCDs, with functional partitioning for power distribution, thermal management, mechanism control, and image data acquisition and transmission to the spacecraft. To validate the core signal chain architecture, particularly the analog-to-digital conversion, FPGA-based processing, and Ethernet data transmission, a dedicated technology demonstrator, AXIS-TAP (ADC Testing and Acquisition Platform), has been designed and is currently in fabrication. AXIS-TAP employs commercial equivalents of radiation-tolerant components (FPGA, ADCs, Ethernet PHY) and interfaces directly with the STA Archon benchtop readout system, enabling simultaneous image acquisition from both systems to validate that the flight-like architecture achieves equivalent or superior noise performance. This paper presents the FEE architectural framework, the AXIS-TAP demonstrator design, and the design principles that enable scalability to future soft X-ray missions with similar detector requirements. Performance validation results will be reported upon completion of AXIS-TAP system integration and testing.

**Keywords:** CCD readout electronics, front-end electronics, FPGA-based data acquisition, soft X-ray detectors

## 1. INTRODUCTION

Advancing time-domain and high-resolution soft X-ray astronomy requires focal plane technology that includes large format, low noise (<3e- rms), fast frame rates (≥5 fps) imaging detectors and modern readout electronics able to withstand the harsh space environment.

Meeting these detector specifications, particularly the need to digitize large numbers of analog signals at high speed while maintaining low-noise performance, presents significant challenges for front-end readout electronics design. Furthermore, the focal plane readout electronics must generate the imaging detector's precision clocks and bias signals and reliably transmit high-rate imaging data to either another data-processing flight computer for on-boarding processing of X-ray events or to downlink them to the ground for offline processing. The Advanced X-ray Imaging Satellite (AXIS)[1][Fig.1] , a NASA Probe-class mission concept, provided the driving requirements and heritage for developing a high-speed camera Front-End Electronics (FEE) architecture to operate a focal plane consisting of 4 Lincoln Lab pJFET CCDs. While AXIS is no longer in active development, the technical solutions developed for its focal plane camera remain relevant and extensible to other missions with similar detector requirements and performance goals.

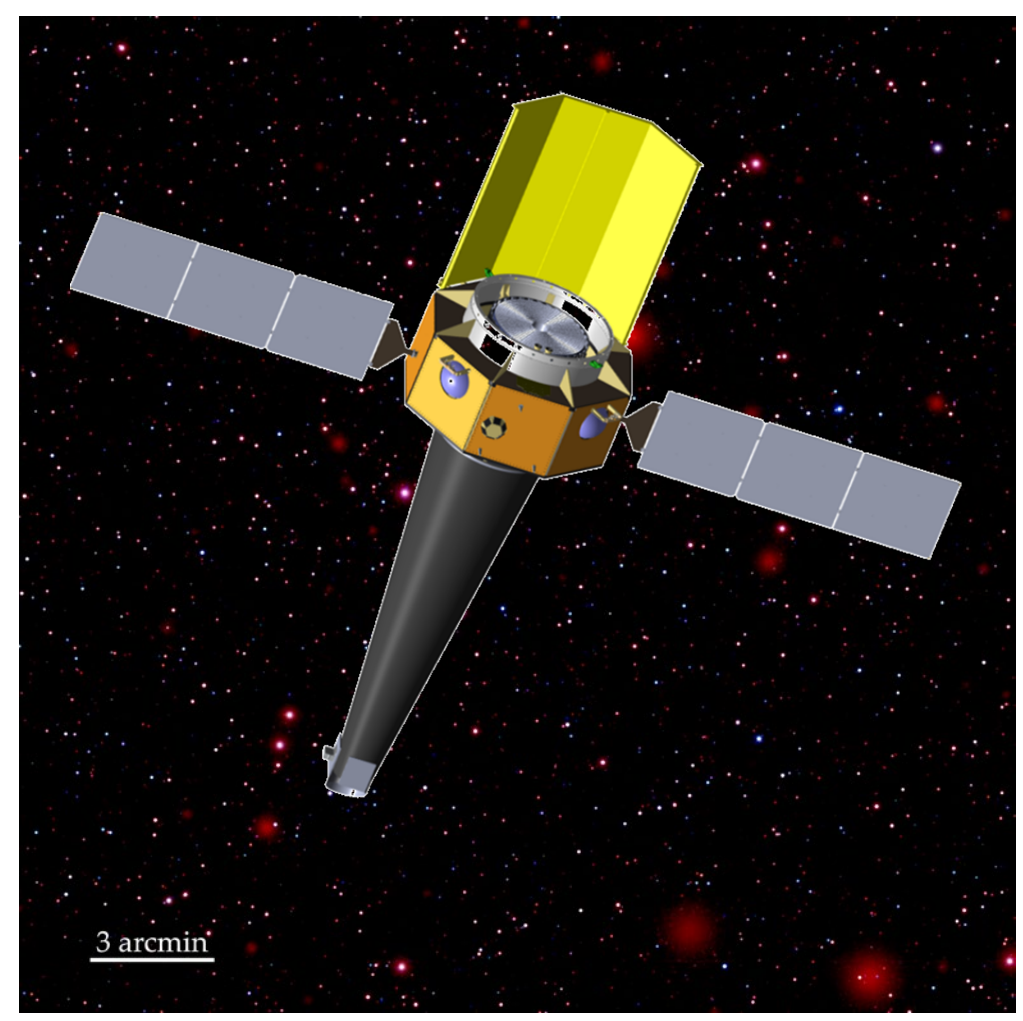

Figure 1: Advanced X-ray Imaging Satellite (AXIS) concept rendering with a simulated observation of an X-ray star field.

This paper presents the FEE design architecture originally developed for AXIS[2][3][4], where the underlying principles can be extended to future soft X-ray payloads. To validate this architecture, we present AXIS-TAP (ADC Testing and Acquisition Platform), a technology demonstrator board currently in fabrication that implements a flight-like frame-grabber architecture (the analog-to-digital conversion, FPGA-based processing, and Ethernet data transmission), using commercial radiation-tolerant equivalents. AXIS-TAP is designed to interface with the bench STA Archon [5] readout system, enabling simultaneous imaging with both systems to validate that the flight-like components meet the noise requirements, acquisition speed, data throughput, and FPGA firmware. The demonstrated ADC chain architecture can extend beyond X-ray missions to other detector types requiring custom analog readout. It is to be noted that the AXIS-TAP demonstrator currently does not provide any clocking or biasing that is necessary to fully operate the detector. The STA Archon controller is still used for this purpose.

## 2. DETECTOR REQUIREMENTS AND PERFORMANCE DRIVERS

For the AXIS mission, the high-speed camera requirements consisted of frame rates at 5 frames per second or above with a read noise of less than 3 $e^-$ rms. The CCID-100 [6][Fig. 2], a custom back-side-illuminated (BSI) pJFET CCD designed by MIT Lincoln Lab is expected to meet these requirements with:

- Format: 1440 x 1455 pixels, 24 x 24 µm pitch
- Output Ports: 16 analog channels
- Full Frame Readout Time: <200 ms at 2 MHz serial rate (enabling > 5fps frame rates)
- Read Noise: < 3 [2.5] e-rms at 3MHz readout
- Quantum Efficiency: >50% QE @ 0.3 keV; >90% @ 1 keV; >90% @ 6 keV
- Energy Resolution: 70eV@0.5keV, 100eV@1keV, 150eV@6 keV
- Detector Thickness: 100 um, operated fully depleted
- Operating Temperature: -110°C (required to achieve low noise specification)
- Buttable Format: 1.6 mm minimum gap on three sides, enabling tight 2×2 focal plane array

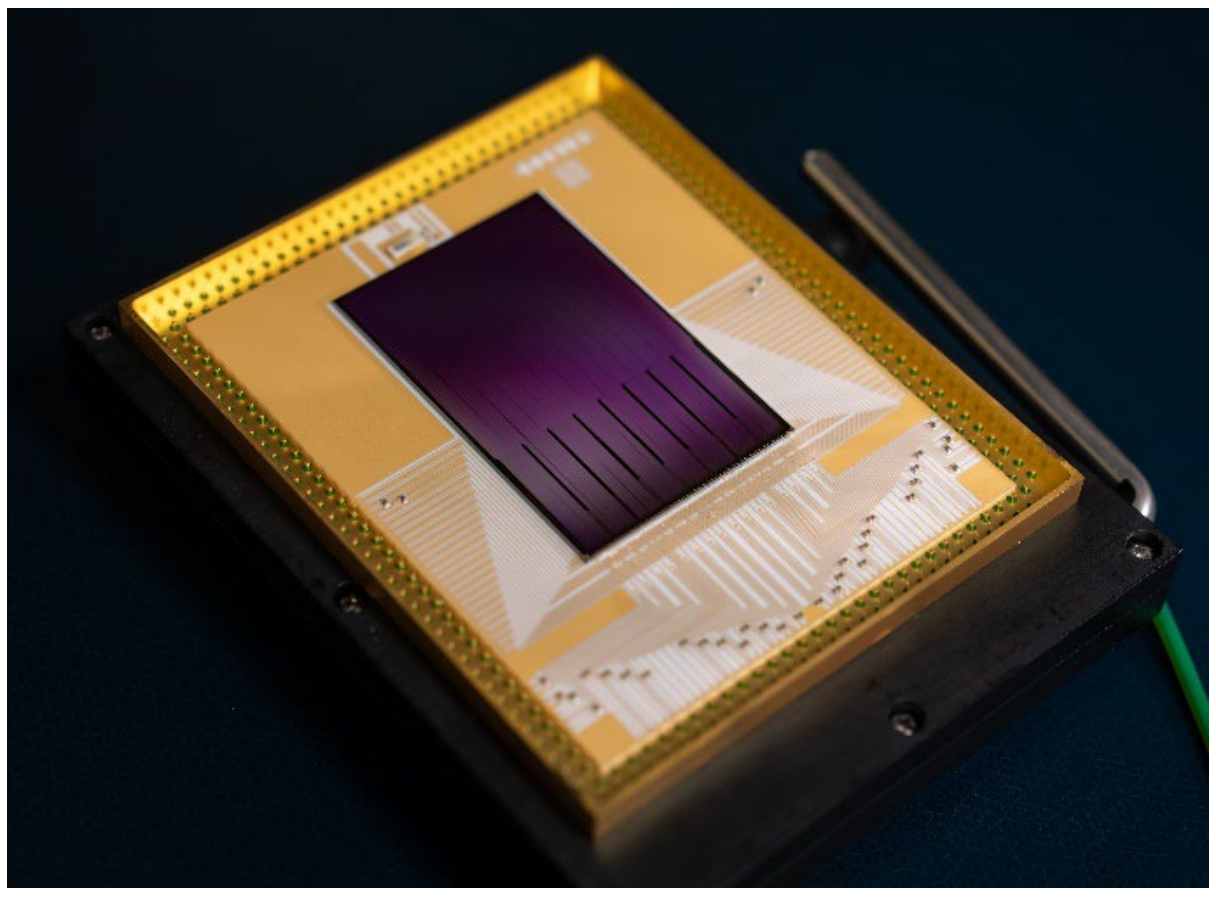

Figure 2: Image of a packaged MIT LL FSI CCID-100, a 16-channel pJFET CCD

## 3. CCID-100 CHARACTERIZATION SETUP

Front-side-illuminated (FSI) CCID-100 devices are currently undergoing characterization, with back-side-illuminated (BSI) variants expected later this year. So far, the results are promising (~3e$^-$ read noise for most outputs sampling at 2MHz serial clocking) [7][8][9]. The CCD devices are delivered in a package socket format from Lincoln Lab. To provide bias voltages, clocking and ADC conversion, an STA Archon, with custom sequencing code specifically written for the CCID-100, is used in conjunction with MIT/Stanford custom designed PCBs including a CCD header board (with a ZIF Socket to hold the CCD), a flex cable with vacuum passthrough, and an Archon interface board. The setup also includes two 8-channel Stanford designed application specific integrated circuit (ASIC), called the Multi-Channel Readout Chip (MCRC)[10]. The ASIC is directly mounted on the CCD header board such to minimize layout parasitics, further, the ASIC provides each channel with its own current source required to bias the detector output stage, a preamplifier to provide gain and an output buffer that can directly drive a long differential 100Ω transmission line. The CCID-100, custom CCD header board and part of the flex cable reside in a custom thermal chamber that enables the device to be operated in a vacuum and cooled to temperatures of -110°C. An X-ray source, such as Fe-55 (Mn Kα ~5.9 keV) can be placed near a Beryllium window on the chamber, or the entire chamber can be mounted onto the MIT IFM Beamline for characterization to lower X-ray energies such as Carbon (277eV) [Fig. 3, Fig. 4].

1455 pixels @ 24um = 34.92 mm

1440 pixels @ 24um = 34.56 mm

Figure 3: An FSI CCID-100 (@-100C) raw frame. Highlighted box includes Mn X-ray events from an Fe-55 source and a cosmic ray event.

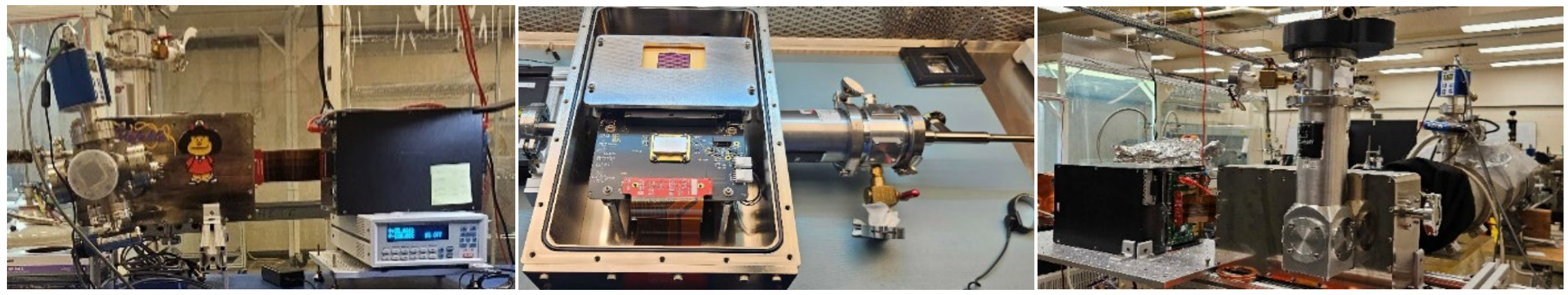

Figure 4: MIT Benchtop characterization setup for the CCID-100 development. (Left) Complete test configuration showing the MIT CCID-100 thermal chamber integrated with the STA Archon readout system. (Middle) Interior view of the thermal chamber showing the CCID-100 sensor, custom header board with Stanford MCRC ASIC preamplifiers, and thermal control components. (Right) CCID-100 test chamber mounted on the MIT IFM Beamline for soft X-ray characterization across multiple photon energies (0.3–1.5 keV and ~6keV).

While the Archon-based setup has enabled initial device characterization, a flight-like readout system is required to validate the complete signal chain architecture and demonstrate performance scalability to the full focal plane array.

## 4. AXIS-TAP TECHNOLOGY DEMONSTRATOR DESIGN

To move towards the development of a flight readout system, a technology demonstration platform, AXIS-TAP [Fig. 5, Fig. 6], has been designed to test critical components of the ADC chain. AXIS-TAP uses commercial chips that have equivalent radiation-tolerant versions, enabling evaluation of the ADC chain, FPGA firmware development, and image generation formatted as UDP packets transmitted over a 1 Gbps Ethernet link.

To enable simultaneous and synchronous imaging through both systems, a custom buffer board mounts [Fig. 6] between the CCD flex cable and the Archon Interface board, using unity gain differential buffers to split the 16 analog signals into two independent paths: one continuing to the Archon, the second to an AXIS-TAP connector via cable.

**Analog Sampling**

A CCD analog signal contains two intervals critical for signal processing: a reset shelf (an interval after sense node reset is complete) and a signal shelf (an interval after signal charge is injected into sense node). Subtracting signal values during these two intervals from each other removes the baseline offset of the pixel. Correlated Double Sampling (CDS) is the traditional approach: sample once on the reset shelf and once on the signal shelf, then subtract the two values. Multi-sampling schemes improve upon this by acquiring multiple samples on each shelf, reducing noise compared to single-sample CDS. This oversampling approach, called Multiple Correlated Double Sampling (MCDS) reduces noise proportional to √(number of samples).

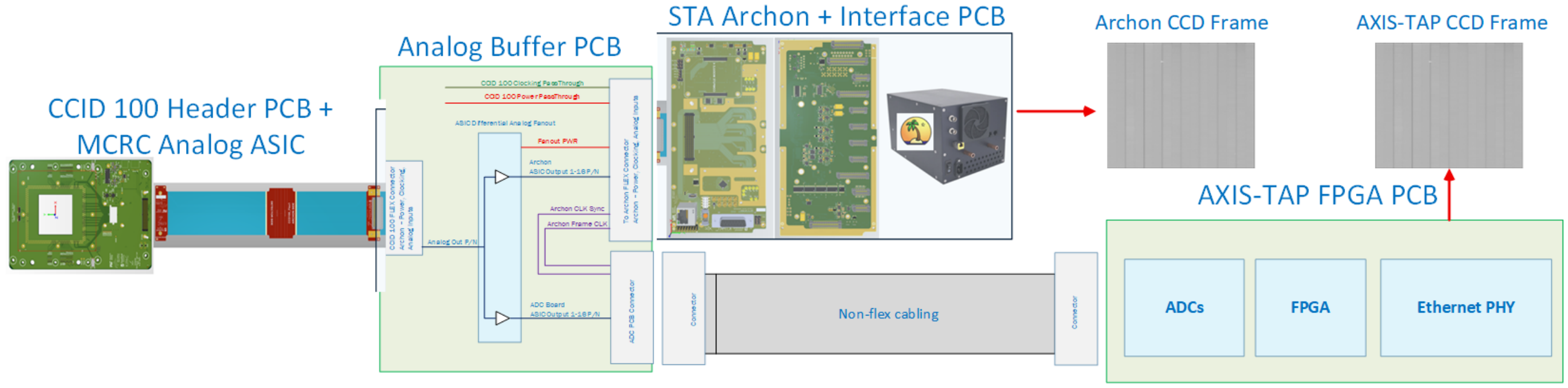

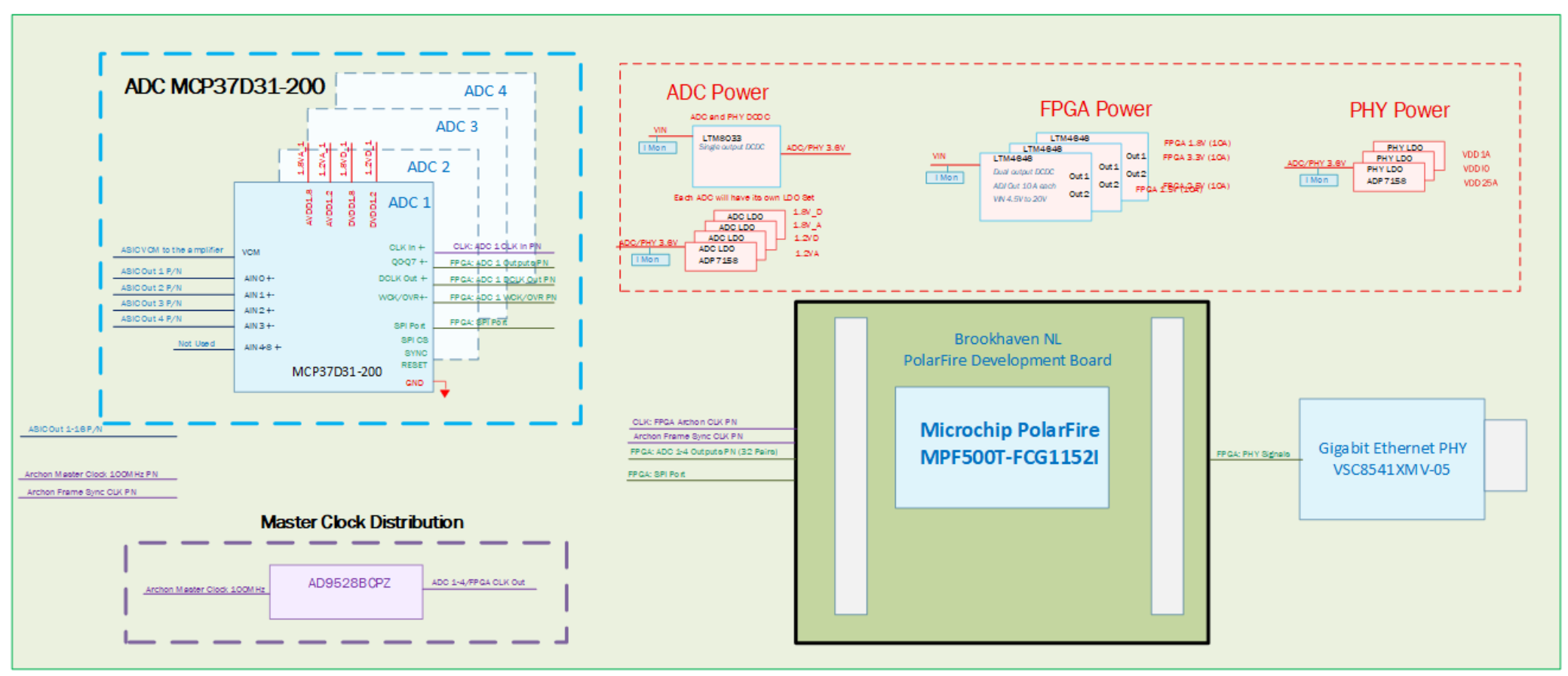

Figure 5: AXIS-TAP system integration with the STA Archon benchtop readout system. (Top) System-level signal flow: The CCID-100 Header PCB with MCRC ASIC (existing) connects via flex cable to the new custom Analog Buffer PCB, which splits the 16 differential analog signals into two parallel paths. One path continues to the existing STA Archon and Interface PCB for Archon-based image acquisition, while the second routes via cabling to the new AXIS-TAP FPGA PCB containing the ADCs, FPGA, and Ethernet PHY for parallel image acquisition. Simultaneous frames from both systems enable direct noise comparison. (Bottom) Detailed view of the AXIS-TAP FPGA PCB.

In the Archon ADC implementation, an MCDS approach is used where multiple samples are acquired with a 10ns clock (50 samples per pixel for the CCID-100 2MHz serial clock). Samples falling within the reset and signal shelves are averaged separately. Note that the samples are not evenly divided between phases: the reset interval transient and the transition between the two shelves are excluded from the averaging. The reset and the signal averages are subtracted to get the final value for the pixel. AXIS-TAP recreates this MCDS scheme: the ADC provides individual samples across the entire pixel signal, and the FPGA firmware performs sample selection (reset vs. signal shelves), averaging, and subtraction.

The selected ADC is the **Microchip MCP37D31-200 (radiation-tolerant equivalent: MCP37D31RT),** a multiplexed 16-bit 200 Msps analog-to-digital converter. While up to 200 Msps can be achieved on a single input, the chip provides 8 inputs with the sampling rate scaled by the number of active inputs. For AXIS-TAP, 4 inputs per chip are used, generating 50 Msps per input. During testing, input selection is programmable, enabling noise scaling studies on individual channels. For the CCID-100 operating at 2 MHz serial clock rate with 50 Msps ADC sampling, the ADC will provide 25 samples per pixel (half the Archon count) if using 4 inputs per chip. Flight implementation will require trade-off analysis between samples per pixel, CCD clocking rate, and the number of ADC chips required to support the full focal plane.

**FPGA**

The FPGA selected is the **Microchip MPF500T-FCG1152I (radiation-tolerant equivalent: RTPF500TS)**, part of Microchip's PolarFire Series, providing flash-based configuration, low power consumption, and sufficient logic capacity, high-speed transceivers, and I/O for the AXIS-TAP application. The FPGA serves as the central processing hub: it controls the four MCP37D31 ADCs, receives high-speed serial sample data, implements the multi-correlated double sampling scheme, and assembles processed pixels into formatted image frames. These frames are packetized as UDP packets and transmitted over a 1 Gbps Ethernet link via a Microchip **VSC8541XMV-05 Ethernet PHY (radiation-tolerant equivalent: VSC8541RT)**. The FPGA additionally handles configuration commands and housekeeping telemetry. The FPGA will also communicate with an RS422 chip, **THVD2452VDR (radiation-tolerant equivalent: THVD9491-SEP)** that could interface to a spacecraft commanding bus.

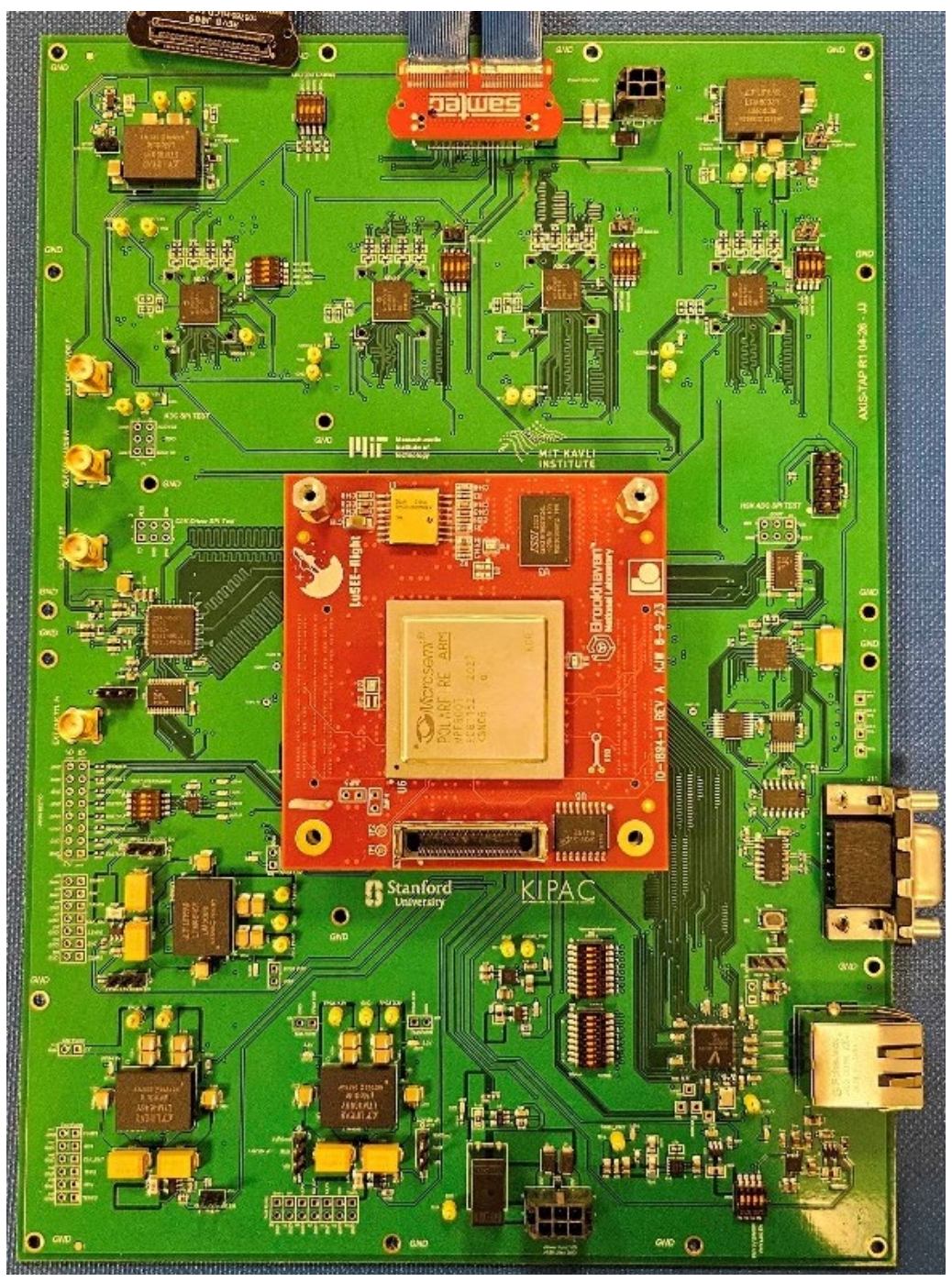

Figure 6: (Top) Buffer PCB that interfaces to the Archon to split the analog signals. (Bottom) AXIS-TAP PCB that includes the ADCs, FPGA and ethernet PHY.

AXIS-TAP implements the FPGA as a mezzanine board designed and developed at Brookhaven National Lab (BNL) [Fig. 7]. The board was developed as a firmware development and prototyping platform for the LuSEE Night spectrometer[11], enabling firmware development and validation on a commercial PolarFire MPF500T FPGA prior to migration to the radiation-tolerant RT PolarFire RTPF500TS used in the flight hardware. Although the MPF500T and RTPF500TS are distinct devices, they are based on the same PolarFire FPGA silicon and differ primarily in packaging and radiation qualification. This common architecture allows firmware developed on the commercial device to be transitioned to the flight FPGA with minimal modifications.

**Power, Clock Distribution and Housekeeping**

The components supporting power distribution, clock generation, and housekeeping monitoring were selected to provide essential support infrastructure and enable rapid design completion, rather than prioritizing radiation-tolerant alternatives. Most of these components leverage heritage from previous validated instrumentation designs.

Power Distribution: FPGA power is distributed via three Analog Devices LTM4646 dual-output DCDC converters (10 A per output), providing low-noise supplies to the FPGA core and I/O rails. ADC power is supplied through an initial voltage conversion stage using an LTM8033 DCDC converter to drop input power to 3.6 V, followed by cascaded low-noise linear regulators (LDOs) for each of the four power rails required by each ADC. **While the LTM8033 lacks a direct radiation-tolerant equivalent, prior screening efforts [12]** indicate it is a candidate for radiation tolerance qualification.

Clock Distribution: A Microchip AD9528 clock distribution IC provides stability and distribution of the 100 MHz pixel clock from the Archon system to the FPGA and four ADCs. The Archon frame synchronization (SYNC) signal connects directly to the FPGA for frame-level timing.

Housekeeping Monitoring: An Analog Devices AD4116 ADC monitors the housekeeping voltages and currents across the board, enabling real-time performance tracking and diagnostics.

**AXIS-TAP: Design Status and Integration**

The AXIS-TAP main board has just completed fabrication and assembly. The BNL FPGA mezzanine boards are currently in-house at MIT and Stanford, respectively. Next steps are firmware development[13], followed by testing and validation against the Archon benchtop system with results to follow in a future paper.

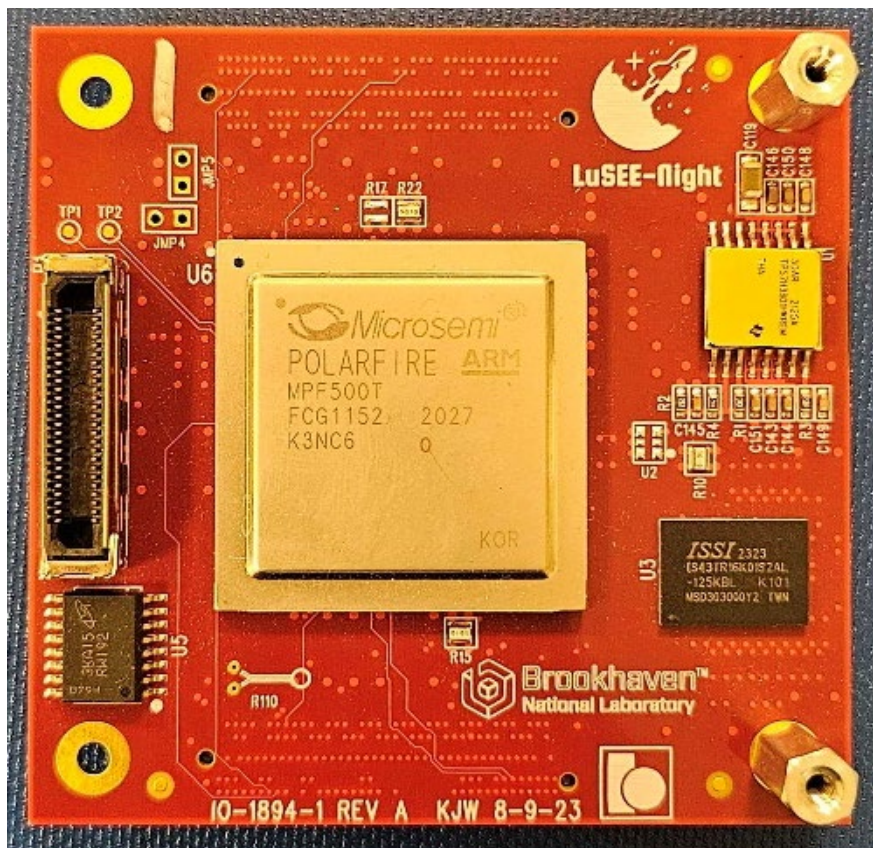

Figure7: BNL developed LuSEE-Night Spectrometer FPGA development PCB used as the FPGA mezzanine in the AXIS-TAP PCB.

## 5. FRONT END ELECTRONICS (FEE) ARCHITECTURE

For the AXIS mission, the intended focal plane [Fig. 8] consists of 4 CCDs that require independent operation, thermal control, contamination control monitoring, and aperture shutter operation. The cameras, heaters, contamination monitor and pressure sensor are to be mounted in a single housing which are vacuum controlled on the ground with a release mechanism to open the detector door in space.

Each CCD has its own MCRC preamplifier ASIC mounted near the CCD. This enables the differential analog signals to be driven out of the camera enclosure to the camera control electronics.

The FEE design is broken into 2 electronic boxes, each with 2 camera control boards and one power distribution unit [Fig. 9]. The camera control board consists of an FPGA dedicated to the CCD operation which drives the bias voltages, clocking and converts the digitized analog signals to frame data. This board also has the ADCs for the incoming CCD analog signals. It collects frames and transmits the data out over a 1 Gb/s ethernet connection. The PDU for each stack provides power conversion and commanding to the camera control boards. The PDU also communicates through an RS422 bus for commanding and housekeeping data.

The FEE receives power, commanding, and transmits raw frames to the Back End electronics (BEE) (Development by Penn State/SWRI). The BEE also performs the spacecraft communication and X-ray event finding which is transmitted to the ground [Fig. 9].

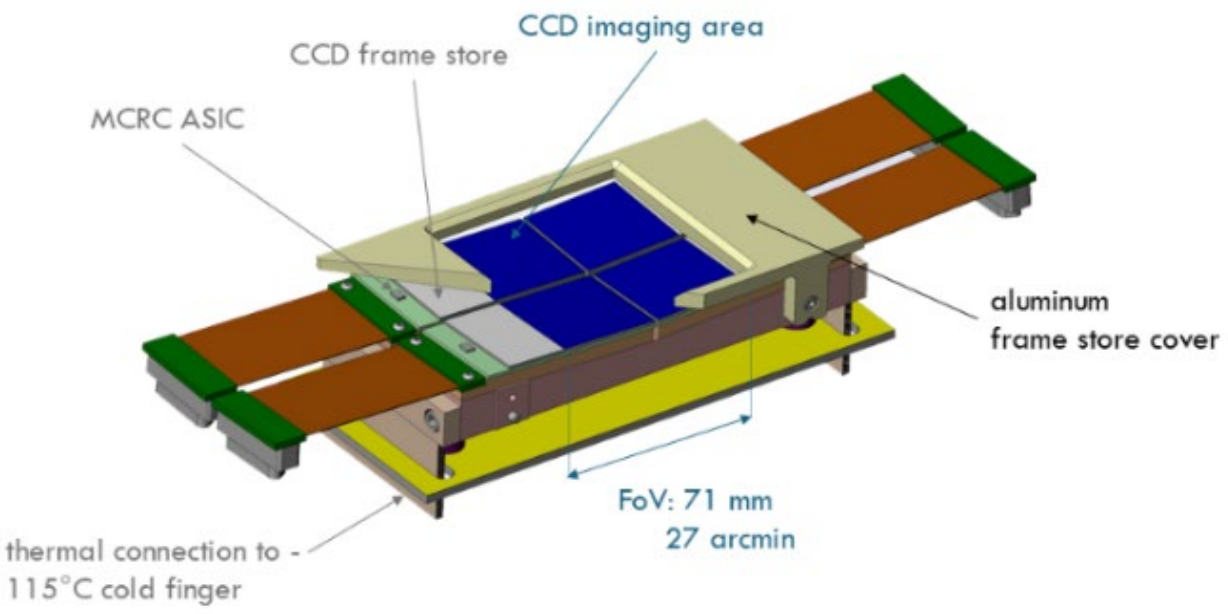

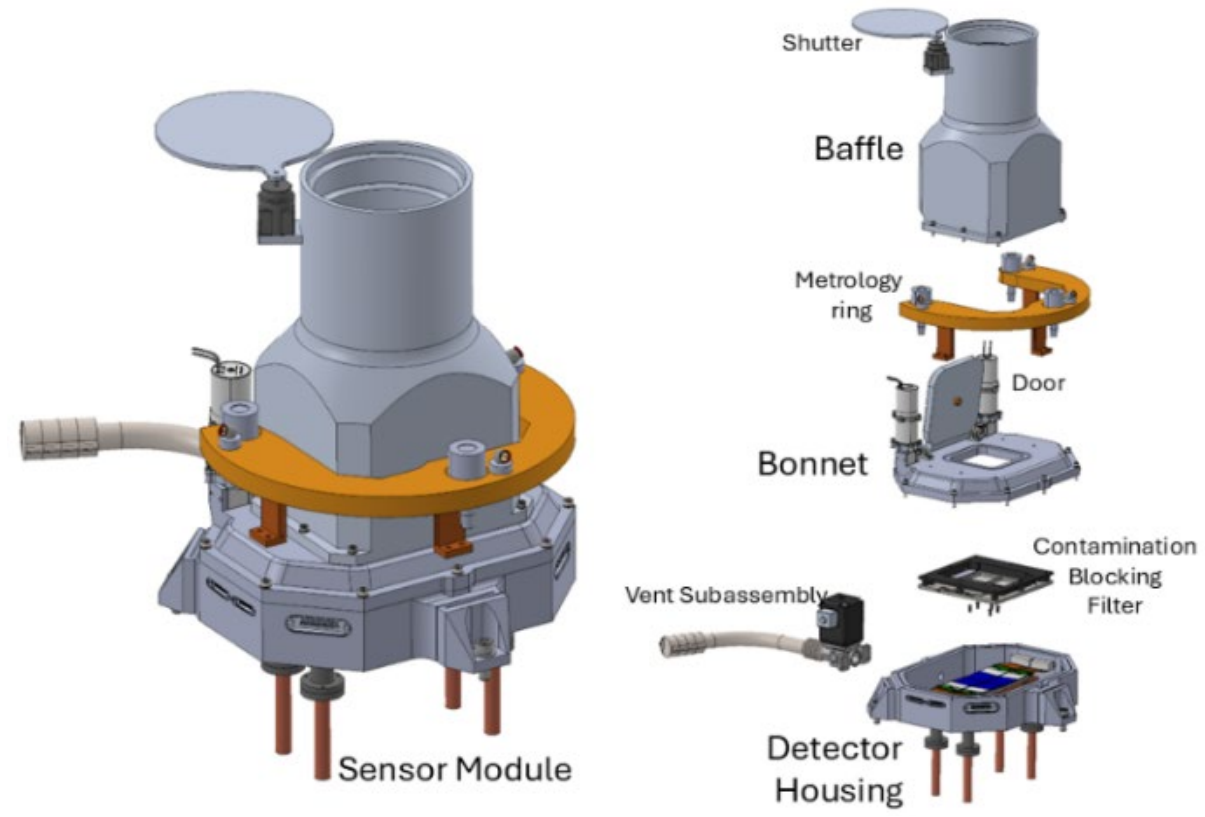

Figure 8: (Top) CAD rendering of the AXIS focal plane showing 4 CCDs butted against each other. (Bottom) CAD rendering of the camera housing showing various components designed for camera operation.

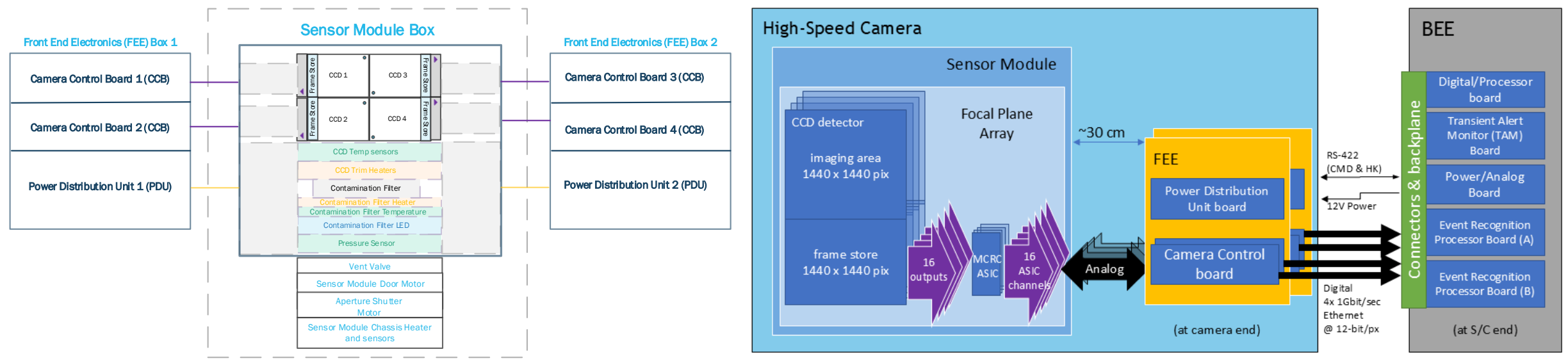

Figure 9: (Left) Block diagram of the main components of the Front End Electronics and Sensor Module. (Right) Data flow of detector data through the FEE to BEE.

## 6. CONCLUSION

Although AXIS is no longer active in development, the technologies presented here for a high-speed camera and front-end electronics can be used in a future X-ray mission. Furthermore, the Phase A portion of the AXIS mission enabled technology development on advancing readout electronics for CCDs in the development of the AXIS-TAP board and ongoing efforts continue for firmware development. Potential advancements include integration of X-ray event finding within the FEE firmware, enabling only identified events to stream downstream rather than raw frames, and adding clock generation and distribution along with other detector control signals like voltage biases.

Furthermore, the ADC chain presented with the AXIS-TAP hardware can potentially be extended to use with other types of detectors such as IR hybridized CMOS sensors that generate analog signal outputs, (e. g. Teledyne HxRG[14], Leonardo LmAPD[15]) as an alternative to commercial detector readout systems (e.g. Archon, Teledyne SIDECAR).

## 7. ACKNOWLEDGEMENTS

We gratefully acknowledge support from NASA for the AXIS Probe Phase A study, under contract 80GSFC25CA019, and from NASA SAT grants 80NNSC23K0211 and 80HQTR24TA005. MIT authors acknowledge additional support from the Kavli Institute for Astrophysics and Space Research, and Stanford authors acknowledge support from the Kavli Institute for Particle Astrophysics and Cosmology.

## REFERENCES

[1] Reynolds, Christopher S., et al. "Overview of the advanced x-ray imaging satellite (AXIS)." *UV, X-Ray, and Gamma-Ray Space Instrumentation for Astronomy XXIII*. Vol. 12678. SPIE, 2023.

[2] Miller, Eric D., et al. "The high-speed x-ray camera on AXIS." *UV, X-Ray, and Gamma-Ray Space Instrumentation for Astronomy XXIII*. Vol. 12678. SPIE, 2023.

[3] Miller, Eric D., et al. "The high-speed x-ray camera on AXIS: design and performance updates." UV, X-Ray, and Gamma-Ray Space Instrumentation for Astronomy XXIV. Vol. 13625. SPIE, 2025.

[4] Grant, Catherine E., et al. "Ground calibration plans for the AXIS high speed camera." *UV, X-Ray, and Gamma-Ray Space Instrumentation for Astronomy XXIV*. Vol. 13625. SPIE, 2025.

[5] STA Instruments, "Archon," [Online]. Available: http://www.sta-inc.net/archon/

[6] Leitz, C., et al. "Development of CCDs for the AXIS astrophysics probe concept." *UV, X-Ray, and Gamma-Ray Space Instrumentation for Astronomy XXIV*. Vol. 13625. SPIE, 2025.

[7] B. J. LaMarr et al., "X-ray characterization of large-format, 16-channel CCD Sensors for future strategic X-ray missions," Proc. SPIE 14157, Paper 14157-92, July 2026.

[8] G. Y. Prigozhin et al., "Noise and charge transfer characteristics of Fast, Low-noise CCD Sensors for AXIS," in Proc. SPIE 14157, Space Telescopes and Instrumentation 2026: Ultraviolet to Gamma Ray, Paper 14157-39, July 2026.
[9] H. R. Stueber et al., "High-speed, low-noise, multi-megapixel CCDs for next generation x-ray observatories," in Proc. SPIE 14157, Space Telescopes and Instrumentation 2026: Ultraviolet to Gamma Ray, Paper 14157-40, July 2026.
[10] Orel, Peter, et al. "X-ray speed reading with the MCRC: prototype success and next generation upgrades." *X-Ray, Optical, and Infrared Detectors for Astronomy XI*. Vol. 13103. SPIE, 2024.
[11] Tamura, Emi, et al. "Design and characterization of the engineering model of the spectrometer onboard LuSEE-Night." *Radio Science* 59.5 (2024): 1-20.
[12] Ameel, Jon, et al. "Radiation-hard power electronics for the ATLAS New Small Wheel." Journal of Instrumentation 10.01 (2015): C01009-C01009.
[13] D. O'Neill et al., "The high-speed FPGA readout system for the Advanced X-ray Imaging Satellite (AXIS)," in Proc. SPIE 14146, Space Telescopes and Instrumentation 2026: Ultraviolet to Gamma Ray, Paper 14146-244, July 2026.
[14] Teledyne Space Imaging, "Infrared Detectors," [Online]. Available: https://www.teledynespaceimaging.com/en-us/products/infrared-detectors
[15] Owton, Daniel, et al. "Leonardo large area avalanche photo-diode arrays." *Infrared Sensors, Devices, and Applications XV*. Vol. 13613. SPIE, 2025.